\documentclass[10pt,a4paper,english]{amsart}
\usepackage[T1]{fontenc}
\usepackage[latin1]{inputenc}
\usepackage{babel}
\usepackage{amsthm}
\usepackage{amstext}
\usepackage{amssymb}
\usepackage{graphicx}
\usepackage{esint}
\usepackage[unicode=true,pdfusetitle,
 bookmarks=true,bookmarksnumbered=false,bookmarksopen=false,
 breaklinks=false,pdfborder={0 0 1},backref=section,colorlinks=false]
 {hyperref}

\makeatletter

\pdfpageheight\paperheight
\pdfpagewidth\paperwidth

\numberwithin{equation}{section}
\numberwithin{figure}{section}
\theoremstyle{plain}
\newtheorem{thm}{\protect\theoremname}
  \theoremstyle{plain}
  \newtheorem{prop}[thm]{\protect\propositionname}
  \theoremstyle{definition}
  \newtheorem{defn}[thm]{\protect\definitionname}
  \theoremstyle{remark}
  \newtheorem{rem}[thm]{\protect\remarkname}

\usepackage{a4wide}
\usepackage{bbm}

\makeatother

  \providecommand{\definitionname}{Definition}
  \providecommand{\propositionname}{Proposition}
  \providecommand{\remarkname}{Remark}
\providecommand{\theoremname}{Theorem}

\begin{document}

\title{Simplicial gauge theory on spacetime}

\author{Tore Gunnar Halvorsen}

\address{Department of Mathematical Sciences\\
Norwegian University of Science and Technology\\
N-7491 Trondheim, Norway}

\email{toregha@gmail.com}

\author{Torquil Macdonald Sørensen}

\address{Centre of Mathematics for Applications\\
University of Oslo\\
P.O.Box 1053 Blindern\\
N-0316 Oslo, Norway}

\email{t.m.sorensen@matnat.uio.no, torquil@gmail.com}
\begin{abstract}
We define a discrete gauge-invariant Yang-Mills-Higgs action on spacetime
simplicial meshes. The formulation is a generalization of classical
lattice gauge theory, and we prove consistency of the action in the
sense of approximation theory. In addition, we perform numerical tests
of convergence towards exact continuum results for several choices
of gauge fields in pure gauge theory.
\end{abstract}


\maketitle
\tableofcontents{}

\section{Introduction}

This article is the natural prolongation of an earlier work \cite{SHC_TGH10}.
In \cite{SHC_TGH10} we defined a gauge invariant discrete action
to approximate the Yang-Mills-Higgs (YMH) action \cite{Yang:1954ek,Weinberg:1995mt,Weinberg:1996kr,Peskin:1995ev}
on a simplicial mesh approximation of a spatial domain, and we proved
consistency of the discrete action for smooth fields. In this article
we expand the discussion to 4d spacetime simplexes in time and three
spatial dimensions, and we also expand the consistency proof to fields
in the natural energy norm.

More precisely, we define the spatial part of the discrete action
as in \cite{SHC_TGH10}, using Whitney forms \cite{whitney57}, Wilson
lines and loops \cite{wilson74,Creutz:1984mg}, and concepts from
FEM \cite{ciarlet78,hiptmair02,monk03}. We then expand the spatial
mesh to spacetime by uniformly discretizing time. The spacetime mesh
is defined by repeating the spatial mesh at every time step. Furthermore,
we extend the Whitney forms to spacetime, yielding a natural expansion
of the discrete spatial action to a spacetime action.

The purpose of this work is to prove consistency of the resulting
approximation scheme for the classical YMH equations, and also to
develop an analogue of lattice gauge theory, i.e. performing quantum
field calculations using the introduced action. The latter will be
described in a companion article \cite{HalvSor2011b}.

We use the same notation as in \cite{SHC_TGH10}, and the article
is organized as follows. In section \ref{sec:YMH_action} we introduce
the continuous YMH action, and also develop the proposed simplicial
gauge theory. In section \ref{sec:diff_exp_map} we review some properties
of the differential of the exponential map for matrix groups and the
Baker-Campbell-Hausdorff formula. We prove consistency of the action
in the energy norm in section \ref{sec:consistency}. Finally, in
section \ref{sec:num_convergence} we describe some numerical convergence
tests, performed on several gauge fields for which the continuum action
was exactly calculable.

\section{Construction\label{sec:YMH_action}}

\subsection{The continuous Yang-Mills action}

Let $\mathbb{M}=\mathbb{R}\times S$ be a riemannian spacetime manifold,
where $\mathbb{R}$ represents time and $S$ is a bounded domain in
three dimensional euclidean space. We let $\mathbb{M}$ be equipped
with a lorentzian or euclidean signature and coordinates $x=(t,\mathbf{x})$.
Furthermore, let $\mathcal{G}$ be a compact Lie group with associated
Lie algebra $\mathfrak{g}$, and assume that $\mathcal{G}$ can be
represented by a subgroup of the complex unitary $n\times n$ matrices,
for some $n$. The hermitian conjugate of a matrix $g$ is denoted
$g^{H}$ and the real valued scalar product is 
\begin{equation}
g'\cdot g:=\Re\text{tr}(g'g^{H}).
\end{equation}

The space of smooth $k$-forms on $\mathbb{M}$ is denoted $\Omega^{k}(\mathbb{M})$,
and the space $\Omega^{k}(\mathbb{M})\otimes\mathfrak{g}$ can be
identified with the space of smooth $\mathfrak{g}$-valued $k$-forms
on $\mathbb{M}$. The bracket of Lie algebra valued forms is defined
as 
\begin{equation}
[u\otimes g,u'\otimes g']:=(u\wedge u')\otimes[g,g'],
\end{equation}
 where $u,u'$ are real valued differential forms and $g,g'\in\mathfrak{g}$.

A connection one-form on $\mathbb{M}$ is an element $\mathbb{A}=(A_{0},A)\in\Omega^{1}(\mathbb{M})\otimes\mathfrak{g}$,
where $A_{0}$ represents the time component and $A$ the spatial
component. The temporal curvature $\mathcal{F}^{t}(\mathbb{A})$ and
spatial curvature $\mathcal{F}^{s}(A)$ of such a one-form are given
by 
\begin{equation}
\mathcal{F}^{t}(\mathbb{A})=dA_{0}+d_{t}A+[A_{0},A],\qquad\mathcal{F}^{s}(A)=dA+\frac{1}{2}[A,A],
\end{equation}
where $d_{t}$ and $d$ denote exterior derivative in the temporal
and spatial direction respectively. We will also use such forms with
lower regularity, see e.g. \cite{christiansen:75}. The independent
variable in Yang-Mills theory is such a connection one-form (not necessarily
smooth), and the action describing it is given by $S[\mathbb{A}]=S[\mathbb{A}]_{t}+S[A]_{s}$,
where 
\begin{equation}
S[\mathbb{A}]_{t}:=\int_{\mathbb{M}}|\mathcal{F}^{t}(\mathbb{A})|^{2},\qquad S[A]_{s}:=\int_{\mathbb{M}}|\mathcal{F}^{s}(A)|^{2}.\label{cont_action}
\end{equation}

A gauge transformation of a connection one-form $\mathbb{A}$ is associated
with a choice of $G\in\mathcal{G}$ for each $x\in\mathbb{M}$, such
that 
\begin{equation}
A_{0}(x)\mapsto G(x)\left(A_{0}(x)+d_{t}\right)G^{-1}(x),\qquad A(x)\mapsto G(x)\left(A(x)+d\right)G^{-1}(x).\label{cont:gauge:trafo}
\end{equation}
The action $S[\mathbb{A}]$ is invariant under such gauge transformations.

\subsection{The interpolated FEM action}

In the FEM formulation, we assume $\mathcal{T}$ to be a simplicial
complex spanning the spatial domain $S$. The simplexes are referred
to as vertexes, edges, faces and tetrahedra according to dimension,
and are labeled $i$, $e$, $f$ and $T$ respectively. The symbol
$T$ will also be used for simplexes of any dimension. We also suppose
an orientation has been chosen for each simplex in $\mathcal{T}$.
In addition, we assume that time is discretized with a time step $\Delta t$,
and that the simplicial complex $\mathcal{T}$ is repeated at every
time step, resulting in a spacetime simplicial complex $\mathbb{T}$.
For a more detailed definition of simplicial complexes, consult \cite[section 5]{christiansen2011}.

Thus, as the basic building block in classical FEM theory is a tetrahedron
$T$, the basic building block in this extended FEM version is $T\times I_{\tau}$,
where $I_{\tau}=[\tau,\tau+\Delta t]$ and $\tau$ denotes the temporal
nodes.

Furthermore, let $W^{k}(\mathcal{T})$ ($W^{k}(T)$) be the space
of Whitney k-forms \cite{whitney57} on $\mathcal{T}$ ($T$), with
canonical basis $(\lambda_{T})$, $T$ ranging over the set of $k$-dimensional
simplexes in $\mathcal{T}$. The 0-forms $\lambda_{i}$ are the barycentric
coordinate maps taking the value 1 at vertex $i$ and 0 at all others.
For an edge $e=\{i,j\}$, the associated Whitney 1-form is defined
by 
\begin{equation}
\lambda_{e}=\lambda_{i}d\lambda_{j}-\lambda_{j}d\lambda_{i},
\end{equation}
and for a face $f=\{i,j,k\}$ the associated Whitney 2-form is defined
by 
\begin{equation}
\lambda_{f}=2\left(\lambda_{i}d\lambda_{j}\wedge d\lambda_{k}+\lambda_{j}d\lambda_{k}\wedge d\lambda_{i}+\lambda_{k}d\lambda_{i}\wedge d\lambda_{j}\right).
\end{equation}
 In order to formulate the Yang-Mills theory in a spacetime FEM setting
we need to extend these k-forms to k-forms on $\mathbb{T}$. In addition,
we need to define the temporal edge and temporal face basis functions,
which are constructed as in \cite{christiansen2011}.

The spatial Whitney k-forms are extended to be piecewise affine in
time and are denoted $(\Lambda_{T(\tau)})$, i.e. 
\begin{equation}
\lambda_{T}\rightarrow\Lambda_{T(\tau)}=\lambda_{T}\otimes P_{1}^{t},
\end{equation}
where $P_{1}^{t}$ denote polynomials in the time variable of degree
at most one, and $T(\tau):=(\tau,T)$ denotes the spatial simplex
$T$ at temporal node $\tau$. More precisely, $\Lambda_{T(\tau)}$
is the piecewise affine function in time, taking the value $\lambda_{T}$
at $\tau$ and 0 at all other temporal nodes. This is consistent with
the requirement that the tangential part of the curvature of the gauge
potential should be continuous across faces.

The temporal edge basis functions are constructed as follows. To every
vertex $i$ in the spatial mesh, there are temporal edges $e_{t}(\tau)=\{i_{\tau},i_{\tau+\Delta t}\}$,
where $i_{\tau}:=i(\tau)$. The temporal basis edge function attached
to $e_{t}(\tau)$ is then the piecewise constant function in time
defined by 
\begin{equation}
\Lambda_{e_{t}(\tau)}(t)=\begin{cases}
\lambda_{i}\circ\pi\frac{1}{\Delta t}dt, & t\in I_{\tau}\\
0, & \text{otherwise}.
\end{cases}
\end{equation}
Here, $\pi$ is the canonical projection onto $S$, i.e. 
\begin{equation}
\pi:\mathbb{M}=\mathbb{R}\times S\rightarrow S,
\end{equation}
and $dt$ is the standard basis one-form in the temporal direction.

Finally, the temporal face elements are constructed as follows. To
every spatial edge $e$ there are corresponding temporal faces $f_{t}(\tau)=e\times I_{\tau}$.
Consider the spatial Whitney edge element $\lambda_{e}$. Then apply
the pull-back of $\pi$ to construct a one-form on spacetime, and
then wedge it with $dt$. More precisely, denote by 
\begin{equation}
\pi^{*}:\Omega(S)\rightarrow\Omega(\mathbb{M}),
\end{equation}
the pull-back induced by $\pi$. Then the temporal basis face function
is 
\begin{equation}
\Lambda_{f_{t}(\tau)}(t)=\begin{cases}
\pi^{*}(\lambda_{e})\wedge\frac{1}{\Delta t}dt=\lambda_{e}\circ\pi\wedge\frac{1}{\Delta t}dt, & t\in I_{\tau}\\
0, & \text{otherwise}.
\end{cases}
\end{equation}

This construction ensures that the temporal face basis is orthogonal
to the spatial face basis. The space spanned by $(\Lambda_{i(\tau)})$,
$(\Lambda_{\mathbbm e(\tau)}):=((\Lambda_{e_{t}(\tau)},\Lambda_{e(\tau)}))$
and $(\Lambda_{\mathbbm f(\tau)}):=((\Lambda_{f_{t}(\tau)},\Lambda_{f(\tau)}))$
are denoted $\mathbb{W}^{0}(\mathbb{T})$, $\mathbb{W}^{1}(\mathbb{T})$
and $\mathbb{W}^{2}(\mathbb{T})$ respectively. If no confusion can
arise, the time dependence index $\tau$ that identifies a temporal
node is usually omitted to compactify notation.

Thus, let $\mathbb{A}=(A_{0},A)\in\mathbb{W}^{1}\otimes\mathfrak{g}$,
$\mathbb{A}=\sum_{e_{t}}A_{0,e_{t}}\Lambda_{e_{t}}+\sum_{e}A_{e}\Lambda_{e}$,
where the summations are over oriented edges, and we remark that 
\begin{equation}
A_{0,e_{t}}=\int_{e_{t}}A_{0},\qquad A_{e}=\int_{e}A.
\end{equation}
The temporal and spatial curvatures of $\mathbb{A}$ are given by
\begin{equation}
\begin{split}\mathcal{F}^{t}(\mathbb{A}) & =\sum_{e}A_{e}d_{t}\Lambda_{e}+\sum_{e_{t}}A_{0,e_{t}}d\Lambda_{e_{t}}+\sum_{e_{t},e}[A_{0,e_{t}},A_{e}]\Lambda_{e_{t}}\wedge\Lambda_{e},\\
\mathcal{F}^{s}(A) & =\sum_{e}A_{e}d\Lambda_{e}+\frac{1}{2}\sum_{e,e'}[A_{e},A_{e'}]\Lambda_{e}\wedge\Lambda_{e'}.
\end{split}
\end{equation}
Since $\Lambda_{\mathbbm e}\wedge\Lambda_{\mathbbm e'}\notin\mathbb{W}^{2}(\mathbb{T})$
we choose to interpolate $\mathcal{F}^{t}(\mathbb{A})$ and $\mathcal{F}^{s}(A)$
onto $\mathbb{W}^{2}(\mathbb{T})$, instead of working with higher
order Whitney elements.

Let $(I^{t},I^{s})$ and $(J^{t},J^{s})$ denote interpolation operators
onto the temporal and spatial Whitney one- and two-forms, respectively.
They are projection operators defined by 
\begin{equation}
\begin{split}I^{t}u & =\sum_{e_{t}}(\int_{e_{t}}u)\Lambda_{e_{t}},\qquad I^{s}u=\sum_{e}(\int_{e}u)\Lambda_{e},\\
J^{t}u & =\sum_{f_{t}}(\int_{f_{t}}u)\Lambda_{f_{t}},\qquad J^{s}u=\sum_{f}(\int_{f}u)\Lambda_{f},
\end{split}
\end{equation}
and are well defined in particular as maps $\Omega^{1}(\mathbb{M})\rightarrow\mathbb{W}^{1}(\mathbb{T})$
and $\Omega^{2}(\mathbb{M})\rightarrow\mathbb{W}^{2}(\mathbb{T})$,
respectively. If we define $I:=I^{t}+I^{s}$, $J:=J^{t}+J^{s}$, and
$\mathbbm d:=(d_{t},d)$, we have $\mathbbm d\circ I=J\circ\mathbbm d$
by Stokes' theorem. In particular, $d\circ I^{s}=J^{s}\circ d$, consistent
with the classical Whitney elements.

The degrees of freedom of the interpolated curvatures are 
\begin{equation}
\begin{split}J_{f_{t}}^{t}(\mathcal{F}^{t}(\mathbb{A})) & :=\int_{f_{t}}\mathcal{F}^{t}(\mathbb{A})=\sum_{e\in f_{t}}A_{e}+\sum_{e_{t}\in f_{t}}A_{0,e_{t}}+\sum_{e,e_{t}\in f_{t}}C_{e_{t}e}[A_{0,e_{t}},A_{e}],\qquad C_{e_{t}e}=\int_{f_{t}}\Lambda_{e_{t}}\wedge\Lambda_{e},\\
J_{f}^{s}(\mathcal{F}^{s}(A)) & :=\int_{f}\mathcal{F}^{s}(A)=\sum_{e\in f}A_{e}+\frac{1}{2}\sum_{e,e'\in f}C_{ee'}[A_{e},A_{e'}],\quad C_{ee'}=\int_{f}\Lambda_{e}\wedge\Lambda_{e'}.
\end{split}
\end{equation}

The interpolated FEM action is then $S^{J}[\mathbb{A}]=S^{J}[\mathbb{A}]_{t}+S^{J}[A]_{s}$,
where 
\begin{equation}
\begin{split}S^{J}[\mathbb{A}]_{t} & =\int_{\mathbb{M}}|J^{t}\mathcal{F}^{t}(\mathbb{A})|^{2}=\\
 & =\Re\sum_{f_{t},f_{t}'}M_{f_{t}f_{t}'}\text{tr}\left(J_{f_{t}}^{t}(\mathcal{F}^{t}(\mathbb{A}))J_{f_{t}'}^{t}(\mathcal{F}^{t}(\mathbb{A}))^{H}\right),\qquad M_{f_{t}f_{t}'}:=\int_{\mathbb{M}}\Lambda_{f_{t}}\cdot\Lambda_{f_{t}'},\\
S^{J}[A]_{s} & =\int_{\mathbb{M}}|J^{s}\mathcal{F}^{s}(A)|^{2}=\\
 & =\Re\sum_{f,f'}M_{ff'}\text{tr}\left(J_{f}^{s}(\mathcal{F}^{s}(A))J_{f'}^{s}(\mathcal{F}^{s}(A))^{H}\right),\qquad M_{ff'}:=\int_{\mathbb{M}}\Lambda_{f}\cdot\Lambda_{f'},
\end{split}
\label{interp_fem_action}
\end{equation}
where $(\cdot)$ denotes the scalar product of alternating forms w.r.t.
the Lorentzian signature.

By inspiration from lattice gauge theory \cite{wilson74,Creutz:1984mg},
we will now in several steps construct an approximation to this action
that is gauge invariant, i.e. invariant under the transformation \eqref{cont:gauge:trafo}.

\subsection{An intermediate action}

\subsubsection*{The spatial part}

Let $\{i,j,k,l\}$ be the vertices of $T$, and pick $A\in W^{1}(T)\otimes\mathfrak{g}$.
Attached to an edge $e=\{i,j\}$ oriented from $i$ to $j$ one has
an element $A_{e}=A_{ij}\in\mathfrak{g}$, where we recall that $A_{e}=\int_{e}A$,
and parallel transport from $i$ to $j$ is given by $U_{ij}=\exp(A_{ij})$.
We suppose $U_{ij}$ to be close enough to 1 so that its logarithm
is unambiguous. We use the sign convention $A_{ij}=-A_{ji}$, which
corresponds to $U_{ji}=U_{ij}^{-1}$.

The discrete spatial curvature associated with a face $f=\{i,j,k\}$
is defined by 
\begin{equation}
F_{ijk}^{s}:=U_{ij}U_{jk}U_{ki}.
\end{equation}
and is in analogy with square Wilson loops in classical lattice gauge
theory \cite{wilson74}. By definition, this formula locates the curvature
at vertex $i$. The curvature at vertex $j$ is related to this by
\begin{equation}
F_{jki}^{s}=U_{ji}F_{ijk}^{s}U_{ij},
\end{equation}
which gives a formula for parallel transport of curvature from $i$
to $j$. Concerning orientation of a face, we notice 
\begin{equation}
F_{ijk}^{s}=(F_{ikj}^{s})^{-1}.
\end{equation}

When $f$ is oriented $i\rightarrow j\rightarrow k$, and the curvature
is located at $i$, we write $F_{f}^{s}=F_{ijk}^{s}$. Thus, we have
defined the spatial curvature of a pointed oriented face $f$. The
distinguished point of $f$ is denoted $\dot{f}$, which in this example
is $\dot{f}=i$.

A discrete gauge transformation is associated with a choice of $G_{i}\in\mathcal{G}$
for each vertex $i$. One then transforms $A$ such that 
\begin{equation}
U_{ij}\mapsto G_{i}U_{ij}G_{j}^{-1},
\end{equation}
implying that the spatial curvature transforms as 
\begin{equation}
F_{f}^{s}\mapsto G_{i}F_{f}^{s}G_{i}^{-1},
\end{equation}
similarly to the gauge transformation of the field strength tensor
in continuous Yang-Mills gauge theory.

\subsubsection*{The temporal part}

Concerning the temporal part of the curvature, the construction is
similar. Let $\{(i_{\tau},j_{\tau},j_{\tau+\Delta t},i_{\tau+\Delta t}\}$
be the vertices of the temporal face $f_{t}(\tau)$, and pick $\mathbb{A}=(A_{0},A)\in\mathbb{W}^{1}\otimes\mathfrak{g}$.
Attached to a spatial edge $e(\tau)=\{i_{\tau},j_{\tau}\}$ one has
as before an element $A_{e(\tau)}=:A_{i_{\tau}j_{\tau}}\in\mathfrak{g}$,
and parallel transport from $i_{\tau}$ to $j_{\tau}$ is again given
by $U_{ij}(\tau)=\exp(A_{i_{\tau}j_{\tau}})$.

Attached to a temporal edge $e_{t}(\tau)=\{i_{\tau},i_{\tau+\Delta t}\}$
one has an element $A_{0,e_{t}(\tau)}=A_{0,i_{\tau}i_{\tau+\Delta t}}\in\mathfrak{g}$,
where we recall $A_{0,e_{t}}=\int_{e_{t}}A_{0}$, and parallel transport
from $i_{\tau}$ to $i_{\tau+\Delta t}$ is given by $U_{0,i_{\tau}i_{\tau+\Delta t}}=\exp(A_{0,i_{\tau}i_{\tau+\Delta t}})$.
We use the sign convention $A_{0,i_{\tau}i_{\tau+\Delta t}}=-A_{0,i_{\tau+\Delta t}i_{\tau}}$
which corresponds to $U_{0,i_{\tau}i_{\tau+\Delta t}}=U_{0,i_{\tau+\Delta t}i_{\tau}}^{-1}$.
Again, we suppose $U_{ij}$ and $U_{0,i_{\tau}i_{\tau+\Delta t}}$
to be close enough to the identity, so that their logarithms are unambiguous.

The discrete temporal curvature associated to $f_{t}(\tau)$ is 
\begin{equation}
F_{i_{\tau},j_{\tau},j_{\tau+\Delta t},i_{\tau+\Delta t}}^{t}=U_{ij}(\tau)U_{0,j_{\tau}j_{\tau+\Delta t}}U_{ji}(\tau+\Delta t)U_{0,i_{\tau+\Delta t}i_{\tau}},
\end{equation}
and again we see the similarity with classical lattice gauge theory.
This formula locates the temporal curvature at vertex $i_{\tau}$.
The curvature at vertex $i_{\tau+\Delta t}$ is 
\begin{equation}
F_{i_{\tau+\Delta t},i_{\tau},j_{\tau},j_{\tau+\Delta t}}^{t}=U_{0,i_{\tau+\Delta t}i_{\tau}}F_{i_{\tau},j_{\tau},j_{\tau+\Delta t},i_{\tau+\Delta t}}^{t}U_{0,i_{\tau}i_{\tau+\Delta t}},
\end{equation}
which gives a formula for parallel transport of curvature from $i_{\tau}$
to $i_{\tau+\Delta t}$. Concerning the orientation of a temporal
face, we notice 
\begin{equation}
F_{i_{\tau},j_{\tau},j_{\tau+\Delta t},i_{\tau+\Delta t}}^{t}=(F_{i_{\tau},i_{\tau+\Delta t},j_{\tau+\Delta t},j_{\tau}}^{t})^{-1}.
\end{equation}

When $f_{t}(\tau)$ is oriented $i_{\tau}\rightarrow j_{\tau}\rightarrow j_{\tau+\Delta t}\rightarrow i_{\tau+\Delta t}$,
and the curvature is located at $i_{\tau}$, we write $F_{f_{t}(\tau)}^{t}=F_{i_{\tau},j_{\tau},j_{\tau+\Delta t},i_{\tau+\Delta t}}^{t}$.
The distinguished point of this pointed oriented face $f_{t}(\tau)$
is denoted $\dot{f}_{t}(\tau)$.

Under a discrete gauge transformation, one transforms $A_{0}$ such
that 
\begin{equation}
U_{0,i_{\tau}i_{\tau+\Delta t}}\mapsto G_{i_{\tau}}U_{0,i_{\tau}i_{\tau+\Delta t}}G_{i_{\tau+\Delta t}}^{-1},
\end{equation}
implying that the temporal curvature transforms as 
\begin{equation}
F_{f_{t}(\tau)}^{t}\mapsto G_{i_{\tau}}F_{f_{t}(\tau)}^{t}G_{i_{\tau}}^{-1}.
\end{equation}

\subsubsection*{The action}

We define the intermediate action as $S^{I}[\mathbb{A}]=S^{I}[\mathbb{A}]_{t}+S^{I}[A]_{s}$,
where 
\begin{equation}
S^{I}[\mathbb{A}]_{t}=\Re\sum_{f_{t},f_{t}'}M_{f_{t}f_{t}'}\text{tr}\left[(F_{f_{t}}^{t}(\mathbb{A})-\mathbbm1)(F_{f_{t}'}^{t}(\mathbb{A})-\mathbbm1)^{H}\right],\label{inter_action_t}
\end{equation}
and 
\begin{equation}
S^{I}[A]_{s}=\Re\sum_{f,f'}M_{ff'}\text{tr}\left[(F_{f}^{s}(A)-\mathbbm1)(F_{f'}^{s}(A)-\mathbbm1)^{H}\right].\label{inter_action_s}
\end{equation}

\subsection{The simplicial gauge theory action}

In this section we use the same notation as in the construction of
the intermediate action, equations \eqref{inter_action_t} and \eqref{inter_action_s}.

The goal is of course to formulate a gauge invariant action. The intermediate
action is not gauge invariant due to interactions between different
faces with non-coincident distinguished points. However, this can
be resolved by parallel transport. We treat the spatial and temporal
part separately.

\subsubsection*{The spatial part}

Let $f$ and $f'$ be two spatial faces of a tetrahedron $T$. The
spatial curvature at $f$ and $f'$ can be connected by the parallel
transport operator $U_{\dot{f}\dot{f}'}$. However, the curvature
associated to the face $f$ at time $\tau$ will interact with the
curvature associated to the face $f'$ not only at time $\tau$, but
also at times $\tau\pm\Delta t$. This is resolved by the parallel
operator in time $U_{0,\dot{f}(\tau)\dot{f}(\tau')}$.

Thus the troubled term 
\begin{equation}
(F_{f(\tau)}^{s}-\mathbbm1)(F_{f'(\tau')}^{s}-\mathbbm1)^{H}
\end{equation}
is replaced by 
\begin{equation}
U_{\dot{f}'(\tau)\dot{f}(\tau)}(F_{f(\tau)}^{s}-\mathbbm1)U_{\dot{f}(\tau)\dot{f}'(\tau)}U_{0,\dot{f}'(\tau)\dot{f}'(\tau')}(F_{f'(\tau')}^{s}-\mathbbm1)^{H}U_{0,\dot{f}'(\tau')\dot{f}'(\tau)},
\end{equation}
and from the transformation properties of $F$ and $U$, its trace
is now gauge invariant.

\subsubsection*{The temporal part}

Let $f_{t}(\tau)$ and $f_{t}'(\tau')$ be two temporal faces. By
properties of the canonical basis $(\Lambda_{f_{t}})$ we know that
interactions between the temporal curvature occur only at equal time
intervals. Thus the term 
\begin{equation}
(F_{f_{t}(\tau)}^{t}-\mathbbm1)(F_{f_{t}'(\tau')}^{t}-\mathbbm1)^{H}
\end{equation}
is replaced by 
\begin{equation}
U_{\dot{f}_{t}'(\tau)\dot{f}_{t}(\tau)}(F_{f_{t}(\tau)}^{t}-\mathbbm1)U_{\dot{f}_{t}(\tau)\dot{f}'_{t}(\tau)}U_{0,\dot{f}_{t}'(\tau)\dot{f}_{t}'(\tau')}(F_{f_{t}'(\tau')}^{t}-\mathbbm1)^{H}U_{0,\dot{f}_{t}'(\tau')\dot{f}_{t}'(\tau)},
\end{equation}
and we note that $\tau=\tau'$ or $\tau=\tau'+\Delta t$. By the transformation
properties of $F$ and $U$, the trace of this term is also gauge
invariant.

\subsubsection*{The action}

We define the simplicial gauge theory (SGT) action as $S^{L}[\mathbb{A}]=S^{L}[\mathbb{A}]_{t}+S^{L}[A]_{s}$
where 
\begin{equation}
\begin{split}S^{L}[\mathbb{A}]_{t}:=\Re\sum_{f_{t}(\tau),f'_{t}(\tau')}M_{f_{t}(\tau),f_{t}'(\tau')}\text{tr} & \Big(U_{\dot{f}'_{t}(\tau)\dot{f}_{t}(\tau)}\left[F_{f_{t}(\tau)}^{t}-\mathbbm1\right]U_{\dot{f}_{t}(\tau)\dot{f}'_{t}(\tau)}\times\\
 & \times U_{0,\dot{f}_{t}'(\tau)\dot{f}'_{t}(\tau')}\left[F_{f_{t}'(\tau')}^{t}-\mathbbm1\right]^{H}U_{0,\dot{f}_{t}'(\tau')\dot{f}'_{t}(\tau)}\Big),
\end{split}
\label{sgt_action}
\end{equation}
and 
\begin{equation}
\begin{split}S^{L}[\mathbb{A}]_{s}:=\Re\sum_{f(\tau),f'(\tau')}M_{f(\tau),f'(\tau')}\text{tr} & \Big(U_{\dot{f}'(\tau)\dot{f}(\tau)}\left[F_{f(\tau)}^{s}-\mathbbm1\right]U_{\dot{f}(\tau)\dot{f}'(\tau)}\times\\
 & \times U_{0,\dot{f}'(\tau)\dot{f}'(\tau')}\left[F_{f'(\tau')}^{s}-\mathbbm1\right]^{H}U_{0,\dot{f}'(\tau')\dot{f}'(\tau)}\Big).
\end{split}
\end{equation}
Thus we can conclude,
\begin{thm}
The simplicial gauge theory action $S^{L}$ is discretely gauge invariant.\label{the:simp_act_gauge_inv}
\end{thm}

\subsection{Gauge invariant scalar field action}

To complete the Yang-Mills-Higgs-action, we add a discrete gauge invariant
action for the scalar Higgs field (however no consistence proof for
it will be provided). The gauge group $\mathcal{G}$ can be represented
by a subgroup of the complex unitary $n\times n$ matrices, and in
that representation the basic scalar fields will form an $n$-tuplet,
i.e. 
\begin{equation}
x\mapsto\phi(x)=\left(\begin{array}{c}
\phi_{1}(x)\\
\vdots\\
\phi_{n}(x)
\end{array}\right)\in\mathbb{C}^{n}.
\end{equation}
The action describing this multi-component field is given by $S[\phi,\mathbb{A}]=S[\phi,A_{0}]_{t}+S[\phi,A]_{s}$,
where 
\begin{equation}
S[\phi,A_{0}]_{t}=\int_{\mathbb{M}}|D_{A_{0}}\phi|^{2},\qquad S[\phi,A]_{s}=\int_{\mathbb{M}}|D_{A}\phi|^{2},
\end{equation}
and $D_{A_{0}}\phi=(d_{t}+A_{0})\phi$ and $D_{A}\phi=(d+A)\phi$
are the gauge-covariant derivatives on scalar fields. We observe that
the action is invariant under the following set of local gauge-transformations
\begin{equation}
\begin{split}\phi(x) & \mapsto G(x)\phi(x),\\
A_{0}(x) & \mapsto G(x)\left(A_{0}(x)+d_{t}\right)G^{-1}(x),\\
A(x) & \mapsto G(x)\left(A(x)+d\right)G^{-1}(x),
\end{split}
\end{equation}
where $G(x)\in\mathcal{G}$. This is so, since the covariant derivatives
transform as $D_{A_{0}}\phi(x)\mapsto G(x)D_{A_{0}}\phi(x)$ and $D_{A}\phi(x)\mapsto G(x)D_{A}\phi(x)$,
and $G^{H}G=\mathbbm1$.

Our aim is to construct a FEM inspired action for this scalar field
that is gauge invariant, and the key ingredient is again the parallel
transport operator $U$ introduced in previous sections.

The scalar field is formally a zero-form, and in FEM it has degrees
of freedom at the nodes of the mesh. Thus, let $\phi\in\mathbb{W}^{0}\otimes\mathbb{C}^{n}$.
Then we can write 
\begin{equation}
\phi=\sum_{i_{\tau}}\phi_{i_{\tau}}\Lambda_{i_{\tau}},\qquad\phi_{i_{\tau}}=\phi(i_{\tau}).
\end{equation}
The covariant derivatives of $\phi$ are one-forms, they have degrees-of-freedom
on the edges of the mesh, and are approximated as in lattice gauge
theory (LGT) \cite{wilson74,Creutz:1984mg}. More precisely, let $e_{t}$
be a temporal edge and $e$ a spatial edge of the mesh. Their origins
are denoted $\dot{e}_{t}$ and $\dot{e}$, and their targets $\ddot{e}_{t}$
and $\ddot{e}$, respectively. Then the components of the covariant
derivative of $\phi$ along these edges are approximated as 
\begin{equation}
\begin{split}(D_{A_{0}}\phi)_{e_{t}} & \approx(\delta_{A_{0}}\phi)_{e_{t}}=\phi_{\ddot{e}_{t}}-U_{0,\ddot{e}_{t}\dot{e}_{t}}\phi_{\dot{e}_{t}},\\
(D_{A}\phi)_{e} & \approx(\delta_{A}\phi)_{e}=\phi_{\ddot{e}}-U_{\ddot{e}\dot{e}}\phi_{\dot{e}}.
\end{split}
\label{disc:cov_derivative}
\end{equation}
 We observe that 
\begin{equation}
(\delta_{A_{0}}\phi)_{e_{t}}\mapsto G_{\ddot{e}_{t}}(\delta_{A_{0}}\phi)_{e_{t}},\qquad(\delta_{A}\phi)_{e}\mapsto G_{\ddot{e}}(\delta_{A}\phi)_{e}
\end{equation}
 whenever 
\begin{equation}
\phi_{i_{\tau}}\mapsto G_{i_{\tau}}\phi_{i_{\tau}},\qquad U_{0,i_{\tau+\Delta t}i_{\tau}}\mapsto G_{i_{\tau+\Delta t}}U_{0,i_{\tau+\Delta t}i_{\tau}}G_{i_{\tau}}^{-1},\qquad U_{ij}\mapsto G_{i}U_{ij}G_{j}^{-1}.
\end{equation}
Thus the components of the approximated covariant derivatives transform
as in the continuous case. However, this is not enough to ensure local
gauge invariance of the action, since the inner product of edge basis
functions involves interactions between different edges.

\subsubsection*{The temporal part}

By using the approximation of the temporal covariant derivative in
equation \eqref{disc:cov_derivative}, a FEM inspired approximation
of $S[\phi,A_{0}]_{t}$ reads 
\begin{equation}
S^{F}[\phi,A_{0}]_{t}=\Re\sum_{e_{t},e_{t}'}M_{e_{t}e_{t}'}(\delta_{A_{0}}\phi)_{e_{t}}^{H}(\delta_{A_{0}}\phi)_{e_{t}'},\qquad M_{e_{t}e_{t}'}=\int_{\mathbb{M}}\Lambda_{e_{t}}\cdot\Lambda_{e_{t}'}.\label{disc:intermed_temporal_scalar_action}
\end{equation}
We note that this approximation is not gauge invariant since the mass
matrix $M_{e_{t}e_{t}'}$ is not diagonal. This problem can be overcome
by either mass-lumping of $M_{e_{t}e_{t}'}$ \cite{christiansen09},
or by using the parallel transport operator $U$ to localize the terms
in the action. The mass-lumping procedure severely restricts the structure
of the mesh for time-dependent problems, thus we choose the second
alternative.

We make the replacement 
\begin{equation}
(\delta_{A_{0}}\phi)_{e_{t}}^{H}(\delta_{A_{0}}\phi)_{e_{t}'}\rightarrow(\delta_{A_{0}}\phi)_{e_{t}}^{H}U_{\ddot{e}_{t}\ddot{e}_{t}'}(\delta_{A_{0}}\phi)_{e_{t}'},
\end{equation}
in equation \eqref{disc:intermed_temporal_scalar_action}, and approximate
the temporal part of the action as 
\begin{equation}
S^{L}[\phi,A_{0}]_{t}=\Re\sum_{e_{t}(\tau),e_{t}'(\tau)}M_{e_{t}(\tau)e_{t}'(\tau)}(\delta_{A_{0}}\phi)_{e_{t}(\tau)}^{H}\, U_{\ddot{e}_{t}(\tau)\ddot{e}_{t}'(\tau)}\,(\delta_{A_{0}}\phi)_{e_{t}'(\tau)}.
\end{equation}
By the transformation properties of $(\delta_{A_{0}}\phi)$ and $U$,
this is gauge-invariant.

\subsubsection*{The spatial part}

By using the approximation of the spatial covariant derivative in
equation \eqref{disc:cov_derivative}, a FEM inspired approximation
of $S[\phi,A]_{s}$ reads 
\begin{equation}
S^{F}[\phi,A]_{s}=\Re\sum_{e,e'}M_{ee'}(\delta_{A}\phi)_{e}^{H}(\delta_{A}\phi)_{e'},\qquad M_{ee'}=\int_{\mathbb{M}}\Lambda_{e}\cdot\Lambda_{e'}.\label{disc:intermed_spatial_scalar_action}
\end{equation}
Again, this approximation is not gauge invariant since the mass matrix
$M_{ee'}$ is not diagonal, and again we choose to resolve this problem
by parallel transport.

We make the replacement 
\begin{equation}
(\delta_{A}\phi)_{e(\tau)}^{H}(\delta_{A}\phi)_{e'(\tau')}\rightarrow(\delta_{A}\phi)_{e(\tau)}^{H}U_{0,\ddot{e}(\tau)\ddot{e}(\tau')}U_{\ddot{e}(\tau')\ddot{e}'(\tau')}\,(\delta_{A}\phi)_{e'(\tau')}
\end{equation}
in equation \eqref{disc:intermed_spatial_scalar_action}, and approximate
the spatial part of the action as 
\begin{equation}
S^{L}[\phi,A]_{s}=\Re\sum_{e(\tau),e'(\tau')}M_{e(\tau)e'(\tau')}(\delta_{A}\phi)_{e(\tau)}^{H}U_{0,\ddot{e}(\tau)\ddot{e}(\tau')}U_{\ddot{e}(\tau')\ddot{e}'(\tau')}(\delta_{A}\phi)_{e'(\tau')}.
\end{equation}
By the transformation properties of $(\delta_{A}\phi)$ and $U$,
this is gauge invariant, We therefore have
\begin{thm}
The action $S^{L}[\phi,\mathbb{A}]=S^{L}[\phi,A_{0}]_{t}+S^{L}[\phi,A]_{s}$
is discretely gauge invariant.
\end{thm}

\section{The differential of the exponential map for matrix groups}

\label{sec:diff_exp_map} Let $\mathcal{G}$ be a compact matrix Lie
group with associated Lie algebra $\mathfrak{g}=T_{\mbox{Id}}\mathcal{G}$.
The structures of $\mathcal{G}$ (the connected component containing
the identity) and $\mathfrak{g}$ are related through the exponential
map 
\begin{equation}
\exp:\mathfrak{g}\rightarrow\mathcal{G},\qquad A\mapsto\exp(A),
\end{equation}
which for matrix Lie groups is given by the usual power series expansion
\begin{equation}
\exp(A)=\sum_{n=0}^{\infty}\frac{A^{n}}{n!}=:e^{A}.
\end{equation}

In order to apply Hamilton's variational principle to the gauge invariant
SGT action introduced above, we need to calculate the differential
of the exponential map in an arbitrary direction. In other words,
we need to find an expression for 
\begin{equation}
D(e^{A})\cdot B=\frac{d}{d\tau}e^{A+tB}\Big|_{\tau=0}.
\end{equation}
If $A$ and $B$ commute, i.e. $[A,B]:=AB-BA=0$, the differential
is straightforward to calculate and equals 
\begin{equation}
D(e^{A})\cdot B=e^{A}B.
\end{equation}
This is not the case if $[A,B]\neq0$.

However, one can prove the following useful formula \cite{hall04}
\begin{prop}
\label{prop:diff_exp} Let $X$ and $Y$ be $n\times n$ ($n\in\mathbb{N}$)
complex matrices. Then the following relation holds 
\begin{equation}
\frac{d}{d\tau}e^{X+\tau Y}\Big|_{\tau=0}=e^{X}\left[\frac{\mathbbm1-e^{-\text{ad}_{X}}}{\text{ad}_{X}}Y\right],
\end{equation}
where $\text{ad}_{X}:\mathfrak{g}\rightarrow\mathfrak{g}$ and $\text{ad}_{X}Y=[X,Y]$.
More generally, if $\tau\mapsto X(\tau)$ is a smooth matrix evaluated
function, then 
\begin{equation}
\frac{d}{d\tau}e^{X(\tau)}=e^{X(\tau)}\left[\frac{\mathbbm1-e^{-\text{ad}_{X(\tau)}}}{\text{ad}_{X(\tau)}}\left(\frac{dX(\tau)}{d\tau}\right)\right].
\end{equation}
 
\end{prop}
By the cyclic invariance of the trace operator, i.e. if $A,B,C$ are
$n\times n$ matrices then $\text{tr}(ABC)=\text{tr}(BCA)=\text{tr}(CAB)$,
one can also prove 
\begin{equation}
\frac{d}{d\tau}\text{tr}(e^{X+\tau Y})\Big|_{\tau=0}=\text{tr}(e^{X}Y).
\end{equation}

\subsection{The Baker-Campbell-Hausdorff formula}

For complex numbers $x$ and $y$ we know that 
\begin{equation}
e^{x}e^{y}=e^{x+y}.
\end{equation}
This is not the case when $x$ and $y$ are replaced by matrices $X$
and $Y$, but the following relation holds \cite{hall04,munthe_owren99}:
\begin{prop}
\label{prop:bch} Let $X$ and $Y$ be $n\times n$ ($n\in\mathbb{N}$)
complex matrices. Then 
\begin{equation}
\begin{split}e^{X}e^{Y} & =e^{Z},\qquad Z=\sum_{n=1}^{\infty}c_{n},\qquad c_{1}=X+Y,\\
(n+1)c_{n+1} & =\frac{1}{2}[X-Y,c_{n}]+\sum_{p=1}^{[n/2]}\frac{B_{2p}}{(2p)!}\sum_{\overset{k_{i}>0,n\geq1}{k_{1}+\ldots+k_{2p}=n}}[c_{k_{1}},[\ldots[c_{k_{2p}},X+Y]\ldots]],
\end{split}
\end{equation}
where $[n/2]$ denotes the largest integer snaller than $n/2$, and $B_{j}$ is the $j$-th
Bernoulli number. 
\end{prop}
The first four terms in the expansion of $Z$ is given by 
\begin{equation}
\begin{split}c_{1} & =X+Y,\qquad c_{2}=\frac{1}{2}(XY-YX)\,,\\
c_{3} & =\frac{1}{12}(X^{2}Y+YX^{2}-2XYX+Y^{2}X+XY^{2}-2YXY)\,,\\
c_{4} & =\frac{1}{24}(X^{2}Y^{2}-2XYXY-Y^{2}X^{2}+2YXYX)\,.
\end{split}
\end{equation}

\section{Consistency\label{sec:consistency} }

Since we have formulated the theory in a spacetime FEM setting, we
choose to define consistency for the entire action, not only for the
spatial part as is usual. This definition of course encompasses the
usual one. Note that we only prove consistency for the gauge field
action, without the scalar field. Inclusion of the scalar field is
a simple extension of this proof.

We suppose that we have a regular sequence of simplicial meshes $\mathcal{T}_{n}$
of the spatial domain $S$. The diameter of a simplex $T$ is denoted
$h_{T}$, and the biggest $h_{T}$ when $T\in\mathcal{T}_{n}$ is
denoted $h_{n}$. In addition, we suppose that time is discretized
by a time step $\Delta t_{n}$, and that $\mathcal{T}_{n}$ is repeated
at every time step, resulting in a simplicial mesh $\mathbb{T}_{n}$
of the spacetime domain $\mathbb{M}$. We suppose that 
\begin{equation}
\max_{\mathbb{T}}\{(h_{n}),(\Delta t_{n})\}\overset{n\rightarrow\infty}{\longrightarrow}0.
\end{equation}

The interpolation operators onto the Whitney elements introduced earlier
are attached with a subscript $n$ to associate them with the mesh
$\mathbb{T}_{n}$. Finally, let $X_{n}=\mathbb{W}^{1}(\mathbb{T}_{n})\otimes\mathfrak{g}$.
\begin{defn}
\label{consistency_def} We say that two actions $S_{n}$ and $S_{n}'$
defined on $X_{n}$ are consistent with each other, with respect to
a norm $\|\cdot\|$, if for any $\mathbb{A}\in X_{n}$ we have 
\begin{equation}
\underset{\mathbb{A}'\in X_{n}}{\sup}\frac{|DS_{n}[\mathbb{A}]\mathbb{A}'-DS'_{n}[\mathbb{A}]\mathbb{A}'|}{\|\mathbb{A}'\|}\underset{n\rightarrow\infty}{\longrightarrow}0.
\end{equation}

\end{defn}
If there is a constant $C$ not depending on $n$ such that quantities
$a_{n}$ and $b_{n}$ satisfy $a_{n}\leq Cb_{n}$ for all $n$, we
write $a_{n}=\mathcal{O}(b_{n})$. To compactify notation the subscript
$n$ will be suppressed.

We have introduced three different actions $S^{J}[\mathbb{A}]$, $S^{I}[\mathbb{A}]$
and $S^{L}[\mathbb{A}]$, and the plan is to show 
\begin{enumerate}
\item $S^{J}[\mathbb{A}]$ consistent with $S[\mathbb{A}]$, 
\item $S^{I}[\mathbb{A}]$ consistent with $S^{J}[\mathbb{A}]$, 
\item $S^{L}[\mathbb{A}]$ consistent with $S^{I}[\mathbb{A}]$, 
\end{enumerate}
which implies the consistency between $S^{L}[\mathbb{A}]$ and $S[\mathbb{A}]$.

We will prove consistency in the energy norm, i.e. 
\begin{equation}
A_{0},A\in L^{\infty}(\mathbb{R};H^{1}(S)),\qquad\partial_{t}A_{0},\partial_{t}A\in L^{\infty}(\mathbb{R};L^{2}(S)),
\end{equation}
where $\partial_{t}$ is a shorthand for $\partial/\partial t$. To
compactify notation we define 
\begin{equation}
\begin{split}\|\cdot\|_{L^{p}(L^{q})} & :=\|\cdot\|_{L^{p}(\mathbb{R};L^{q}(S))},\quad\forall\,0<p,q\leq\infty,\\
\|\cdot\|_{L^{\infty}(H^{1})} & :=\|\cdot\|_{L^{\infty}(\mathbb{R};H^{1}(S))},\\
\|\mathbb{A}\| & :=\|A_{0}\|+\|A\|.
\end{split}
\end{equation}
The $H^{1}$ spacetime Euclidean seminorm is denoted $|\cdot|_{H^{1}(\mathbb{M})}$.

\subsection{Consistency between $S^{J}[\mathbb{A}]$ and $S[\mathbb{A}]$}

The two actions are given in equations \eqref{interp_fem_action}
and \eqref{cont_action} respectively, and 
\begin{equation}
\begin{split}|DS[\mathbb{A}]\cdot\mathbb{A}'-DS^{J}[\mathbb{A}]\cdot\mathbb{A}'|\leq K_{t}+K_{s},\\
K_{t}:=|DS[\mathbb{A}]_{t}\cdot\mathbb{A}'-DS^{J}[\mathbb{A}]_{t}\cdot\mathbb{A}'|,\\
K_{s}:=|DS[A]_{s}\cdot A'-DS^{J}[A]_{s}\cdot A'|.
\end{split}
\end{equation}
We treat $K_{t}$ and $K_{s}$ separately. To compactify notation
we define 
\begin{equation}
\begin{split}\mathbbm d_{\mathbb{A}}\mathbb{A}' & :=dA_{0}'+d_{t}A'+[A_{0}',A]+[A_{0},A'],\\
d_{A}A' & :=dA'+[A,A'],\\
\partial A & :=\partial_{t}A+\nabla A,\\
\partial A_{0} & :=\partial_{t}A_{0}+\nabla A_{0},
\end{split}
\end{equation}
where $\nabla$ is the spatial gradient on k-forms.

We first estimate the term $K_{t}$. From equations \eqref{interp_fem_action}
and \eqref{cont_action} we get 
\begin{equation}
\begin{split}K_{t} & =2\Big|\int_{\mathbb{M}}\Big(\langle\mathcal{F}^{t}(\mathbb{A}),\mathbbm d_{\mathbb{A}}\mathbb{A}'\rangle-\langle J^{t}\mathcal{F}^{t}(\mathbb{A}),J^{t}\mathbbm d_{\mathbb{A}}\mathbb{A}'\rangle\Big)\Big|\\
 & \leq2\|\mathcal{F}^{t}(\mathbb{A})\|_{L^{2}(L^{2})}\|\mathbbm d_{\mathbb{A}}\mathbb{A}'-J^{t}\mathbbm d_{\mathbb{A}}\mathbb{A}'\|_{L^{2}(L^{2})}+2\|\mathcal{F}^{t}(\mathbb{A})-J^{t}\mathcal{F}^{t}(\mathbb{A})\|_{L^{2}(L^{2})}\|J^{t}\mathbbm d_{\mathbb{A}}\mathbb{A}'\|_{L^{2}(L^{2})}.
\end{split}
\end{equation}
The interpolation operator $J^{t}$ possesses two important properties.
First of all it is a projection operator, and second it is stable
$L^{2}\rightarrow L^{2}$ by scaling. This implies that 
\begin{equation}
\begin{split}\|\mathbbm d_{\mathbb{A}}\mathbb{A}'- & J^{t}\mathbbm d_{\mathbb{A}}\mathbb{A}'\|_{L^{2}(L^{2})}\leq Ch\left(|[A_{0}',A]|_{H^{1}(\mathbb{M})}+|[A_{0},A']|_{H^{1}(\mathbb{M})}\right)\\
 & \leq Ch\left(\|A_{0}'\partial A\|_{L^{2}(L^{2})}+\|A\partial A_{0}'\|_{L^{2}(L^{2})}+\|A'\partial A_{0}\|_{L^{2}(L^{2})}+\|A_{0}\partial A'\|_{L^{2}(L^{2})}\right)\\
 & \leq Ch\big(\|A_{0}'\|_{L^{\infty}(L^{6})}\|\partial A\|_{L^{\infty}(L^{3})}+\|A\|_{L^{\infty}(L^{6})}\|\partial A_{0}'\|_{L^{\infty}(L^{3})}+\\
 & \qquad\qquad\qquad+\|A'\|_{L^{\infty}(L^{6})}\|\partial A_{0}\|_{L^{\infty}(L^{3})}+\|A_{0}\|_{L^{\infty}(L^{6})}\|\partial A'\|_{L^{\infty}(L^{3})}\big)\\
 & \leq Ch^{1/2}\big(\|A_{0}'\|_{L^{\infty}(L^{6})}\|\partial A\|_{L^{\infty}(L^{2})}+\|A\|_{L^{\infty}(L^{6})}\|\partial A_{0}'\|_{L^{\infty}(L^{2})}+\\
 & \qquad\qquad\qquad+\|A'\|_{L^{\infty}(L^{6})}\|\partial A_{0}\|_{L^{\infty}(L^{2})}+\|A_{0}\|_{L^{\infty}(L^{6})}\|\partial A'\|_{L^{\infty}(L^{2})}\big)\\
 & \leq Ch^{1/2}\left(\|\mathbb{A}'\|_{L^{\infty}(H^{1})}+\|\partial_{t}\mathbb{A}'\|_{L^{\infty}(L^{2})}\right)
\end{split}
\end{equation}
 where $C$ is a generic constant, $\|\partial\mathbb{A}'\|_{L^{\infty}(L^{2})}:=\|\partial A_{0}'\|_{L^{\infty}(L^{2})}+\|\partial A'\|_{L^{\infty}(L^{2})}$
and $\|\partial_{t}\mathbb{A}'\|_{L^{\infty}(L^{2})}:=\|\partial_{t}A_{0}'\|_{L^{\infty}(L^{2})}+\|\partial_{t}A'\|_{L^{\infty}(L^{2})}$.
By similar arguments 
\begin{equation}
\|\mathcal{F}^{t}(\mathbb{A})-J^{t}\mathcal{F}^{t}(\mathbb{A})\|_{L^{2}(L^{2})}\leq Ch^{1/2}\left(\|A_{0}\|_{L^{\infty}(L^{6})}\|\partial A\|_{L^{\infty}(L^{2})}+\|A\|_{L^{\infty}(L^{6})}\|\partial A_{0}\|_{L^{\infty}(L^{2})}\right).
\end{equation}
Combining the above estimates together with the stability of $J^{t}$
we get 
\begin{equation}
K_{t}\leq Ch^{1/2}\left(\|\mathbb{A}'\|_{L^{\infty}(H^{1})}+\|\partial_{t}\mathbb{A}'\|_{L^{\infty}(L^{2})}\right).
\end{equation}

\begin{rem}
If $\mathbb{A}$ is smooth, then 
\begin{equation}
K_{t}\leq Ch\left(\|\mathbb{A}'\|_{L^{\infty}(H^{1})}+\|\partial_{t}\mathbb{A}'\|_{L^{\infty}(L^{2})}\right).
\end{equation}
 
\end{rem}
From equations \eqref{interp_fem_action} and \eqref{cont_action},
we estimate the term $K_{s}$ as 
\begin{equation}
\begin{split}K_{s} & =2\Big|\int_{\mathbb{M}}\Big(\langle\mathcal{F}^{s}(A),d_{A}A'\rangle-\langle J^{s}\mathcal{F}^{s}(A),J^{s}d_{A}A'\rangle\Big)\Big|\\
 & \leq2\|\mathcal{F}^{s}(A)\|_{L^{2}(L^{2})}\|d_{A}A'-J^{s}d_{A}A'\|_{L^{2}(L^{2})}+2\|\mathcal{F}^{s}(A)-J^{s}\mathcal{F}^{s}(A)\|_{L^{2}(L^{2})}\|J^{s}d_{A}A'\|_{L^{2}(L^{2})}.
\end{split}
\end{equation}

\begin{rem}
By similar arguments as we used to bound $K_{t}$ we get 
\begin{equation}
K_{s}\leq Ch^{1/2}\left(\|A'\|_{L^{\infty}(H^{1})}+\|\partial_{t}A'\|_{L^{\infty}(L^{2})}\right).
\end{equation}
If $A$ is smooth, then 
\begin{equation}
K_{s}\leq Ch\left(\|A'\|_{L^{\infty}(H^{1})}+\|\partial_{t}A'\|_{L^{\infty}(L^{2})}\right).
\end{equation}

\end{rem}
Summing up we get 
\begin{equation}
|DS[\mathbb{A}]\cdot\mathbb{A}'-DS^{J}[\mathbb{A}]\cdot\mathbb{A}'|\leq C\left(\|\mathbb{A}'\|_{L^{\infty}(H^{1})}+\|\partial_{t}\mathbb{A}'\|_{L^{\infty}(L^{2})}\right)\begin{cases}
h^{1/2} & \mathbb{A}\in L^{\infty}(H^{1}),\partial_{t}\mathbb{A}\in L^{\infty}(L^{2})\\
h & \mathbb{A}\quad\text{smooth},
\end{cases}
\end{equation}
implying consistency between $S[\mathbb{A}]$ and $S^{J}[\mathbb{A}]$.

\subsection{Consistency between $S^{I}[\mathbb{A}]$ and $S^{J}[\mathbb{A}]$}

The interpolated FEM action is given in equation \eqref{interp_fem_action},
and the differential of it is 
\begin{equation}
\begin{split}DS^{J}[\mathbb{A}]\cdot\mathbb{A}' & =DS^{J}[\mathbb{A}]_{t}\cdot\mathbb{A}'+DS^{J}[A]_{s}\cdot A',\\
DS^{J}[\mathbb{A}]_{t}\cdot\mathbb{A}' & =\Re\sum_{f_{t},f'_{t}}M_{f_{t}f'_{t}}\text{tr}(J_{f_{t}}^{t}(\mathcal{F}^{t}(\mathbb{A})J_{f_{t}'}^{t}(\mathbbm d_{\mathbb{A}}\mathbb{A}')^{H}+\text{h.c.}),\\
DS^{J}[A]_{s}\cdot A' & =\Re\sum_{f,f'}M_{ff'}\text{tr}(J_{f}^{s}(\mathcal{F}^{s}(A))J_{f'}^{s}(d_{A}A')^{H}+\text{h.c.}).
\end{split}
\end{equation}

Let $f_{t}=\{i_{\tau},j_{\tau},j_{\tau+\Delta t},i_{\tau+\Delta t}\}$
be a temporal face, oriented from $i_{\tau}\rightarrow j_{\tau}\rightarrow j_{\tau+\Delta t}\rightarrow i_{\tau+\Delta t}$,
and write $e_{1}=\{(i_{\tau},j_{\tau}\}$, $e_{2}=\{j_{\tau},j_{\tau+\Delta t}\}$,
$e_{3}=\{j_{\tau+\Delta t},i_{\tau+\Delta t}\}$, and $e_{4}=\{i_{\tau+\Delta t},i_{\tau}\}$.
For such a face, the constants $C_{e_{t}e}=-C_{ee_{t}}$ in the definition
of $J^{t}$ take the value $C_{e_{1}e_{2}}=C_{e_{2}e_{3}}=C_{e_{3}e_{4}}=C_{e_{4}e_{1}}=\frac{1}{4}$.
Hence 
\begin{equation}
\begin{split} & J_{f_{t}}^{t}(\mathbbm d_{\mathbb{A}}\mathbb{A}')=A_{e_{1}}'+A_{0,e_{2}}'+A_{e_{3}}'+A_{0,e_{4}}'+\frac{1}{4}([A'_{e_{1}},A_{0,e_{2}}]+[A_{e_{1}},A'_{0,e_{2}}])+\\
 & \frac{1}{4}([A'_{0,e_{2}},A_{e_{3}}]+[A_{0,e_{2}},A'_{e_{3}}])+\frac{1}{4}([A'_{e_{3}},A_{0,e_{4}}]+[A_{e_{3}},A'_{0,e_{4}}])+\frac{1}{4}([A'_{0,e_{4}},A_{e_{1}}]+[A_{0,e_{4}},A'_{e_{1}}]).
\end{split}
\end{equation}
In addition, we know that $A_{e_{1}}+A_{e_{3}}=(d_{t}A)_{f_{t}}$
and $A_{0,e_{2}}+A_{0,e_{4}}=(dA_{0})_{f_{t}}$, implying that $A_{e_{3}}=-A_{e_{1}}+(d_{t}A)_{f_{t}}$
and $A_{0,e_{4}}=-A_{0,e_{2}}+(dA_{0})_{f_{t}}$, with similar relations
for $A_{e_{3}}'$ and $A_{0,e_{4}}'$. This gives 
\begin{equation}
\begin{split}J_{f_{t}}^{t}(\mathbbm d_{\mathbb{A}}\mathbb{A}') & =A_{e_{1}}'+A_{0,e_{2}}'+A_{e_{3}}'+A_{0,e_{4}}'+([A_{e_{1}}',A_{0,e_{2}}]+[A_{e_{1}},A_{0,e_{2}}'])+\\
 & \qquad\qquad+\mathcal{O}(A_{0}d_{t}A'+AdA_{0}'+\mathbb{A}\leftrightarrow\mathbb{A}')_{f_{t}},
\end{split}
\end{equation}
where e.g. $\mathcal{O}(A_{0}d_{t}A')_{f_{t}}$ means $\mathcal{O}((A_{0})_{e_{t}}(d_{t}A')_{f_{t}})$,
$e_{t}\in f_{t}$.

Let $f=\{i,j,k\}$ be a spatial face, oriented from $i\rightarrow j\rightarrow k$,
and write $e_{1}=\{i,j\}$, $e_{2}=\{j,k\}$, $e_{3}=\{k,i\}$. For
such a face, the constants $C_{ee'}=-C_{e'e}$ in the definition of
$J^{s}$ take the value $C_{e_{1}e_{2}}=C_{e_{3}e_{1}}=C_{e_{2}e_{3}}=\frac{1}{6}$.
Hence 
\begin{equation}
\begin{split}J_{f}^{s}(d_{A}A') & =A_{e_{1}}'+A_{e_{2}}'+A_{e_{3}}'+\frac{1}{6}([A'_{e_{1}},A_{e_{2}}]+[A_{e_{1}},A'_{e_{2}}])+\\
 & \qquad\frac{1}{6}([A'_{e_{1}},A_{e_{3}}]+[A_{e_{1}},A'_{e_{3}}])+\frac{1}{6}([A'_{e_{2}},A_{e_{3}}]+[A_{e_{2}},A'_{e_{3}}]).
\end{split}
\end{equation}
In addition, we know that $A_{e_{1}}+A_{e_{2}}+A_{e_{3}}=(dA)_{f}$,
which means that $A_{e_{3}}=-A_{e_{1}}-A_{e_{2}}+(dA)_{f}$, with
a similar relation for $A'_{e_{3}}$, implying that 
\begin{equation}
J_{f}^{s}(d_{A}A')=A_{e_{1}}'+A_{e_{2}}'+A_{e_{3}}'+\frac{1}{2}([A'_{e_{1}},A_{e_{2}}]+[A_{e_{1}},A'_{e_{2}}])+\mathcal{O}(AdA'+A'dA)_{f}.
\end{equation}
Again, $\mathcal{O}(AdA')_{f}=\mathcal{O}(A_{e}(dA')_{f})$, $e\in f$.

Furthermore we need to calculate the differential of $S^{I}[A]$.
The intermediate action is given in equations \eqref{inter_action_t}
and \eqref{inter_action_s}, and the differential of it is 
\begin{equation}
\begin{split}DS^{I}[\mathbb{A}]\cdot\mathbb{A}' & =DS^{I}[\mathbb{A}]_{t}\cdot\mathbb{A}'+DS^{I}[A]_{s}\cdot A',\\
DS^{I}[\mathbb{A}]_{t}\cdot\mathbb{A}' & =\Re\sum_{f_{t},f'_{t}}M_{f_{t}f'_{t}}\text{tr}\left(\frac{dF_{f_{t}}^{t}(\mathbb{A}+\tau\mathbb{A}')}{d\tau}\Big|_{\tau=0}(F_{f'_{t}}^{t}(\mathbb{A})-\mathbbm1)^{H}+\text{h.c.}\right),\\
DS^{I}[A]_{s}\cdot A' & =\Re\sum_{f,f'}M_{ff'}\text{tr}\left(\frac{dF_{f}^{s}(A+\tau A')}{d\tau}\Big|_{\tau=0}(F_{f'}^{s}(A)-\mathbbm1)^{H}+\text{h.c.}\right).
\end{split}
\end{equation}
Thus we need to calculate 
\begin{equation}
\frac{dF_{f_{t}}^{t}(\mathbb{A}+\tau\mathbb{A}')}{d\tau}\Big|_{\tau=0},\qquad\frac{dF_{f}^{s}(A+\tau A')}{d\tau}\Big|_{\tau=0}.
\end{equation}
We choose to calculate this by first writing the exponential functions
in $F$ as a single exponential using the BCH formula, and then using
the formula for the differential, proposition \ref{prop:diff_exp}.

Let $f_{t}$ be as above. Then $F_{f_{t}}^{t}(\mathbb{A})=U_{e_{1}}U_{e_{2}}U_{e_{3}}U_{e_{4}}$,
where $U_{e_{1}}=\exp(A_{e_{1}})$ and with similar expressions for
$U_{e_{2}}$, $U_{e_{3}}$ and $U_{e_{4}}$. By the BCH formula, proposition
\ref{prop:bch}, $F_{f_{t}}^{t}(\mathbb{A}+\tau\mathbb{A}')$ can
be written as 
\begin{equation}
F_{f_{t}}^{t}(\mathbb{A}+\tau\mathbb{A}')=e^{A_{e_{1}}(\tau)}e^{A_{0,e_{2}}(\tau)}e^{A_{e_{3}}(\tau)}e^{A_{0,e_{4}}(\tau)}=e^{W_{f_{t}}(A_{e_{1}},A_{0,e_{2}},A_{e_{3}},A_{0,e_{4}})(\tau)},
\end{equation}
where $\mathbb{A}(\tau)=\mathbbm A+\tau\mathbbm A'$, and $W_{f_{t}}(A_{e_{1}},A_{0,e_{2}},A_{e_{3}},A_{0,e_{4}})(\tau)$
is given from the recursion formula in proposition \ref{prop:bch}.
If we write $W_{f_{t}}(\tau):=W_{f_{t}}(A_{e_{1}},A_{0,e_{2}},A_{e_{3}},A_{0,e_{4}})(\tau)=\sum_{n=1}^{\infty}d_{n}$,
then the two first terms are 
\begin{equation}
\begin{split}d_{1} & =A_{e_{1}}(\tau)+A_{0,e_{2}}(\tau)+A_{e_{3}}(\tau)+A_{0,e_{4}}(\tau),\\
d_{2} & =\frac{1}{2}[A_{e_{1}}(\tau),A_{0,e_{2}}(\tau)]+\frac{1}{2}[A_{e_{1}}(\tau),A_{e_{3}}(\tau)]+\frac{1}{2}[A_{e_{1}}(\tau),A_{0,e_{4}}(\tau)]+\\
 & \qquad\qquad+\frac{1}{2}[A_{0,e_{2}}(\tau),A_{e_{3}}(\tau)]+\frac{1}{2}[A_{0,e_{2}}(\tau),A_{0,e_{4}}(\tau)]+\frac{1}{2}[A_{e_{3}}(\tau),A_{0,e_{4}}(\tau)].
\end{split}
\end{equation}
We note from proposition \ref{prop:bch} that $d_{n}\propto\mathcal{O}(\frac{1}{n}\mathbb{A}(\tau)^{n})$.
If we again use $A_{e_{3}}=-A_{e_{1}}+(d_{t}A)_{f_{t}}$ and $A_{0,e_{4}}=-A_{0,e_{2}}+(dA_{0})_{f_{t}}$,
with similar relations for $A_{e_{3}}'$ and $A_{0,e_{4}}'$, then
we get 
\begin{equation}
\begin{split}W_{f_{t}}(\tau=0) & =J_{f_{t}}^{t}(\mathcal{F}^{t}(\mathbb{A}))+\mathcal{O}(\mathbb{A}^{3}+\mathbb{A}\mathbbm d\mathbb{A})_{f_{t}},\\
\frac{dW_{f_{t}}(\tau)}{d\tau}\Big|_{\tau=0} & =J_{f_{t}}^{t}(\mathbbm d_{\mathbb{A}}\mathbb{A}')+\mathcal{O}(\mathbb{A}^{2}\mathbb{A}'+\mathbb{A}\mathbbm d\mathbb{A}'+\mathbb{A}'\mathbbm d\mathbb{A})_{f_{t}},
\end{split}
\end{equation}
which again implies that 
\begin{equation}
F_{f_{t}}^{t}(\mathbb{A})=e^{W_{f_{t}}(\tau=0)}=\mathbbm1+J_{f_{t}}^{t}(\mathcal{F}^{t}(\mathbb{A}))+\mathcal{O}(\mathbb{A}^{3}+\mathbb{A}\mathbbm d\mathbb{A})_{f_{t}}.
\end{equation}

Using proposition \ref{prop:diff_exp}, we get 
\begin{equation}
\frac{dF_{f_{t}}^{t}(\mathbb{A}+\tau\mathbb{A}')}{d\tau}\Big|_{\tau=0}=F_{f_{t}}^{t}(\mathbb{A})\left[\frac{\mathbbm1-e^{-ad_{W_{f_{t}}(0)}}}{ad_{W_{f_{t}}(0)}}\,\frac{dW_{f_{t}}(\tau)}{d\tau}\Big|_{\tau=0}\right],
\end{equation}
where 
\begin{equation}
\frac{\mathbbm1-e^{-ad_{W_{f}(0)}}}{ad_{W_{f}(0)}}=\mathbbm1-\frac{1}{2}ad_{W_{f}(0)}+\frac{1}{3!}ad_{W_{f}(0)}^{2}+\ldots,
\end{equation}
and where we recall that $ad_{X}Y=[X,Y]$.

Finally, combining the above estimates gives 
\begin{equation}
\begin{split}(F_{f_{t}'}(\mathbb{A})-\mathbbm1)^{H} & F_{f_{t}}(\mathbb{A})\left[\frac{\mathbbm1-e^{-ad_{W_{f_{t}}(0)}}}{ad_{W_{f_{t}}(0)}}\,\frac{dW_{f_{t}}(\tau)}{d\tau}\Big|_{\tau=0}\right]=\\
 & =J_{f_{t}'}^{t}(\mathcal{F}^{t}(\mathbb{A}))^{H}\frac{dW_{f_{t}}(\tau)}{d\tau}\Big|_{\tau=0}+\mathcal{O}(\mathbb{A}^{4}\mathbb{A}')_{f_{t}f'_{t}}\\
 & =J_{f_{t}'}^{t}(\mathcal{F}^{t}(\mathbb{A}))^{H}J_{f_{t}}^{t}(\mathbbm d_{\mathbb{A}}\mathbb{A}')+\mathcal{O}(\mathbb{A}^{4}\mathbb{A}'+(\mathbb{A}^{3}+\mathbb{A}\mathbbm d\mathbb{A})\mathbbm d\mathbb{A}'+(\mathbb{A}^{2}\mathbbm d\mathbb{A}+(\mathbbm d\mathbb{A})^{2})\mathbb{A}')_{f_{t}f'_{t}}.
\end{split}
\end{equation}

Let $f$ be as above. Then $F_{f}^{s}(A)=U_{e_{1}}U_{e_{2}}U_{e_{3}}$,
where $U_{e_{1}}=\exp(A_{e_{1}})$ and with similar expressions for
$U_{e_{2}}$ and $U_{e_{3}}$. By the BCH formula, proposition \ref{prop:bch},
$F_{f}^{s}(A)$ can be written as 
\begin{equation}
F_{f}^{s}(A+\tau A')=e^{A_{e_{1}}(\tau)}e^{A_{e_{2}}(\tau)}e^{A_{e_{3}}(\tau)}=e^{W_{f}(A_{e_{1}},A_{e_{2}},A_{e_{3}})(\tau)},
\end{equation}
where $A(\tau)=A+\tau A'$, and $W_{f}(A_{e_{1}},A_{e_{2}},A_{e_{3}})(\tau)$
is given from the recursion formula in proposition \ref{prop:bch}.
If we write $W_{f}(\tau):=W_{f}(A_{e_{1}},A_{e_{2}},A_{e_{3}})(\tau)=\sum_{n=1}^{\infty}d_{n}$,
then the first two terms are 
\begin{equation}
\begin{split}d_{1}= & A_{e_{1}}(\tau)+A_{e_{2}}(\tau)+A_{e_{3}}(\tau),\\
d_{2}= & \frac{1}{2}[A_{e_{1}}(\tau),A_{e_{2}}(\tau)]+\frac{1}{2}[A_{e_{1}}(\tau),A_{e_{3}}(\tau)]+\frac{1}{2}[A_{e_{2}}(\tau),A_{e_{3}}(\tau)],
\end{split}
\end{equation}
and we note that $d_{n}\propto\mathcal{O}(\frac{1}{n}A(\tau)^{n})$.
If we again use that $A_{e_{3}}=-A_{e_{1}}-A_{e_{2}}+(dA)_{f}$, with
a similar relation for $A'_{e_{3}}$, then we get 
\begin{equation}
\begin{split}W_{f}(\tau=0) & =J_{f}^{s}(\mathcal{F}^{s}(A))+\mathcal{O}(A^{3}+AdA)_{f},\\
\frac{dW_{f}(t)}{dt}\Big|_{t=0} & =J_{f}^{s}(d_{A}A')+\mathcal{O}(A^{2}A'+AdA'+A'dA)_{f},
\end{split}
\end{equation}
which again implies that 
\begin{equation}
F_{f}^{s}(A)=e^{W_{f}(t=0)}=\mathbbm1+J_{f}^{s}(\mathcal{F}^{s}(A))+\mathcal{O}(A^{3}+AdA)_{f}.
\end{equation}

Using proposition \ref{prop:diff_exp}, we get 
\begin{equation}
\frac{dF_{f}^{s}(A+\tau A')}{d\tau}\Big|_{\tau=0}=F_{f}^{s}(A)\left[\frac{\mathbbm1-e^{-ad_{W_{f}(0)}}}{ad_{W_{f}(0)}}\,\frac{dW_{f}(\tau)}{d\tau}\Big|_{\tau=0}\right],
\end{equation}
and by combining the above estimates gives 
\begin{equation}
\begin{split}(F_{f'}^{s}(A)-\mathbbm1)^{H} & F_{f}^{s}(A)\left[\frac{\mathbbm1-e^{-ad_{W_{f}(0)}}}{ad_{W_{f}(0)}}\,\frac{dW_{f}(\tau)}{d\tau}\Big|_{\tau=0}\right]=\\
 & =J_{f'}^{s}(\mathcal{F}^{s}(A))^{H}\frac{dW_{f}(\tau)}{d\tau}\Big|_{\tau=0}+\mathcal{O}(A^{4}A')_{ff'}\\
 & =J_{f'}^{s}(\mathcal{F}^{s}(A))^{H}J_{f}^{s}(d_{A}A')\\
 & \quad+\mathcal{O}(A^{4}A'+(A^{3}+AdA)dA'+(A^{2}dA+(dA)^{2})A')_{ff'}.
\end{split}
\end{equation}
Summing up, we get 
\begin{equation}
\begin{split}DS^{I}[\mathbb{A}]_{t}\cdot\mathbb{A}' & =\Re\sum_{f_{t},f_{t}'}M_{f_{t}f_{t}'}\text{tr}\left((F_{f_{t}'}^{t}(\mathbb{A})-\mathbbm1)^{H}F_{f_{t}}^{t}(\mathbb{A})\left[\frac{\mathbbm1-e^{-ad_{W_{f_{t}}(0)}}}{ad_{W_{f_{t}}(0)}}\,\frac{dW_{f_{t}}(\tau)}{d\tau}\Big|_{\tau=0}\right]+\text{h.c.}\right)\\
 & =\Re\sum_{f_{t},f_{t}'}M_{f_{t}f_{t}'}\text{tr}\left(J_{f'_{t}}^{t}(\mathcal{F}^{t}(\mathbb{A}))^{H}J_{f_{t}}^{t}(\mathbbm d_{\mathbb{A}}\mathbb{A}')+\text{h.c.}\right)+\\
 & \qquad+\Re\sum_{f_{t},f_{t}'}M_{f_{t}f_{t}'}\text{tr}\left(\mathcal{O}\left(\mathbb{A}^{4}\mathbb{A}'+(\mathbb{A}^{3}+\mathbb{A}\mathbbm d\mathbb{A})\mathbbm d\mathbb{A}'+(\mathbb{A}^{2}\mathbbm d\mathbb{A}+(\mathbbm d\mathbb{A})^{2})\mathbb{A}'\right)_{f_{t}f_{t}'}\right)\\
 & =DS^{J}[\mathbb{A}]_{t}\cdot\mathbb{A}'+\\
 & \qquad+\Re\sum_{f_{t},f_{t}'}M_{f_{t}f_{t}'}\text{tr}\left(\mathcal{O}\left(\mathbb{A}^{4}\mathbb{A}'+(\mathbb{A}^{3}+\mathbb{A}\mathbbm d\mathbb{A})\mathbbm d\mathbb{A}'+(\mathbb{A}^{2}\mathbbm d\mathbb{A}+(\mathbbm d\mathbb{A})^{2})\mathbb{A}'\right)_{f_{t}f_{t}'}\right).
\end{split}
\end{equation}
and 
\begin{equation}
\begin{split}DS^{I}[A]_{s}\cdot A' & =\Re\sum_{f,f'}M_{ff'}\text{tr}\left((F_{f'}(A)-\mathbbm1)^{H}F_{f}(A)\left[\frac{\mathbbm1-e^{-ad_{W_{f}(0)}}}{ad_{W_{f}(0)}}\,\frac{dW_{f}(\tau)}{d\tau}\Big|_{\tau=0}\right]+\text{h.c.}\right)\\
 & =\Re\sum_{f,f'}M_{ff'}\text{tr}\left(J_{f'}^{s}(\mathcal{F}^{s}(A))^{H}J_{f}^{s}(d_{A}A')+\text{h.c.}\right)+\\
 & \qquad+\Re\sum_{f,f'}M_{ff'}\text{tr}\left(\mathcal{O}\left(A^{4}A'+(A^{3}+AdA)dA'+(A^{2}dA+(dA)^{2})A'\right)_{ff'}\right)\\
 & =DS^{J}[A]_{s}\cdot A'+\\
 & \qquad+\Re\sum_{f,f'}M_{ff'}\text{tr}\left(\mathcal{O}\left(A^{4}A'+(A^{3}+AdA)dA'+(A^{2}dA+(dA)^{2})A'\right)_{ff'}\right).
\end{split}
\end{equation}
If we assume that the mesh satisfies a CFL condition, i.e. there exists
a constant $C$ such that 
\begin{equation}
0<\frac{1}{C}\leq\frac{\Delta t_{n}}{h_{n}}\leq C,\qquad\forall n,
\end{equation}
then we can deduce the bounds 
\begin{equation}
\begin{split}|A_{e}| & \leq Ch^{1/2}\|A\|_{L^{\infty}(L^{6})},\qquad|A_{0,e_{t}}|\leq Ch^{1/2}\|A_{0}\|_{L^{\infty}(L^{6})},\\
|(dA)_{f}| & \leq Ch^{1/2}\|\partial A\|_{L^{\infty}(L^{2})},\qquad|(d_{t}A)_{f_{t}}|\leq Ch^{1/2}\|\partial A\|_{L^{\infty}(L^{2})},\\
|(dA_{0})_{f_{t}}| & \leq Ch^{1/2}\|\partial A_{0}\|_{L^{\infty}(L^{2})},
\end{split}
\end{equation}
and we can conclude 
\begin{equation}
|DS^{J}[\mathbb{A}]\cdot\mathbb{A}'-DS^{I}[\mathbb{A}]\cdot\mathbb{A}'|\leq C\left(\|\mathbb{A}'\|_{L^{\infty}(H^{1})}+\|\partial_{t}\mathbb{A}'\|_{L^{\infty}(L^{2})}\right)\begin{cases}
h^{1/2} & \mathbb{A}\in L^{\infty}(H^{1}),\partial_{t}\mathbb{A}\in L^{\infty}(L^{2})\\
h & \mathbb{A}\quad\text{smooth},
\end{cases}\label{cons_fem_inter}
\end{equation}
implying consistency between $S^{J}[\mathbb{A}]$ and $S^{I}[\mathbb{A}]$.

\subsection{Consistency between $S^{I}[\mathbb{A}]$ and $S^{L}[\mathbb{A}]$}

The only difference between $S^{I}$ and $S^{L}$ is the parallel
transport operators in $S^{L}$ introduced to make $S^{L}$ gauge
invariant. They are given as 
\begin{equation}
\begin{split}U_{\dot{f}\dot{f}'} & =\exp(A_{\dot{f}\dot{f}'})=\mathbbm1+A_{\dot{f}\dot{f}'}+\mathcal{O}(A^{2}),\\
U_{0,\dot{f}(\tau)\dot{f}(\tau')} & =\exp(A_{0,\dot{f}(\tau)\dot{f}(\tau')})=\mathbbm1+A_{0,\dot{f}(\tau)\dot{f}(\tau')}+\mathcal{O}(A_{0}^{2}),
\end{split}
\end{equation}
and the differential of these are proportional to $A'$ and $A_{0}'$
respectively. Hence, by similar considerations as in the previous
section we get exactly the same estimate as in equation \eqref{cons_fem_inter},
i.e. 
\begin{equation}
|DS^{I}[\mathbb{A}]\cdot\mathbb{A}'-DS^{L}[\mathbb{A}]\cdot\mathbb{A}'|\leq C\left(\|\mathbb{A}'\|_{L^{\infty}(H^{1})}+\|\partial_{t}\mathbb{A}'\|_{L^{\infty}(L^{2})}\right)\begin{cases}
h^{1/2} & \mathbb{A}\in L^{\infty}(H^{1}),\partial_{t}\mathbb{A}\in L^{\infty}(L^{2})\\
h & \mathbb{A}\quad\text{smooth}.
\end{cases}
\end{equation}
We summarize the results in a theorem: 
\begin{thm}
Assume $\mathbb{M}$ is a bounded domain in $\mathbb{R}^{1+3}$. Then
the SGT action \eqref{sgt_action}, is consistent with the continuous
Yang-Mills action \eqref{cont_action}, with respect to the norm 
\begin{equation}
\|\mathbb{A}\|:=\|\mathbb{A}\|_{L^{\infty}(H^{1})}+\|\partial_{t}\mathbb{A}\|_{L^{\infty}(L^{2})},
\end{equation}
under the assumption that the above-mentioned CFL condition holds.
\end{thm}
As a consequence of the above estimates, we get the following estimate
for the deviation of the SGT action $S^{L}$ from the continuous action
$S$,

\begin{equation}
|S(\mathbb{A})-S^{L}(\mathbb{A})|\leq\begin{cases}
Ch & ,\mathbb{A}\in L^{\infty}(H^{1}),\partial_{t}\mathbb{A}\in L^{\infty}(L^{2})\\
Ch^{2} & ,\mathbb{A}\quad\mbox{smooth}.
\end{cases}\label{eq:action-error}
\end{equation}

\section{Numerical convergence tests\label{sec:num_convergence}}

The preceeding sections have defined and proven consistency of the
SGT action. However, due to the complexity involved in these quantities,
we would like to include some numerical convergence tests as well.

\begin{figure}

\begin{centering}
\includegraphics[width=7cm]{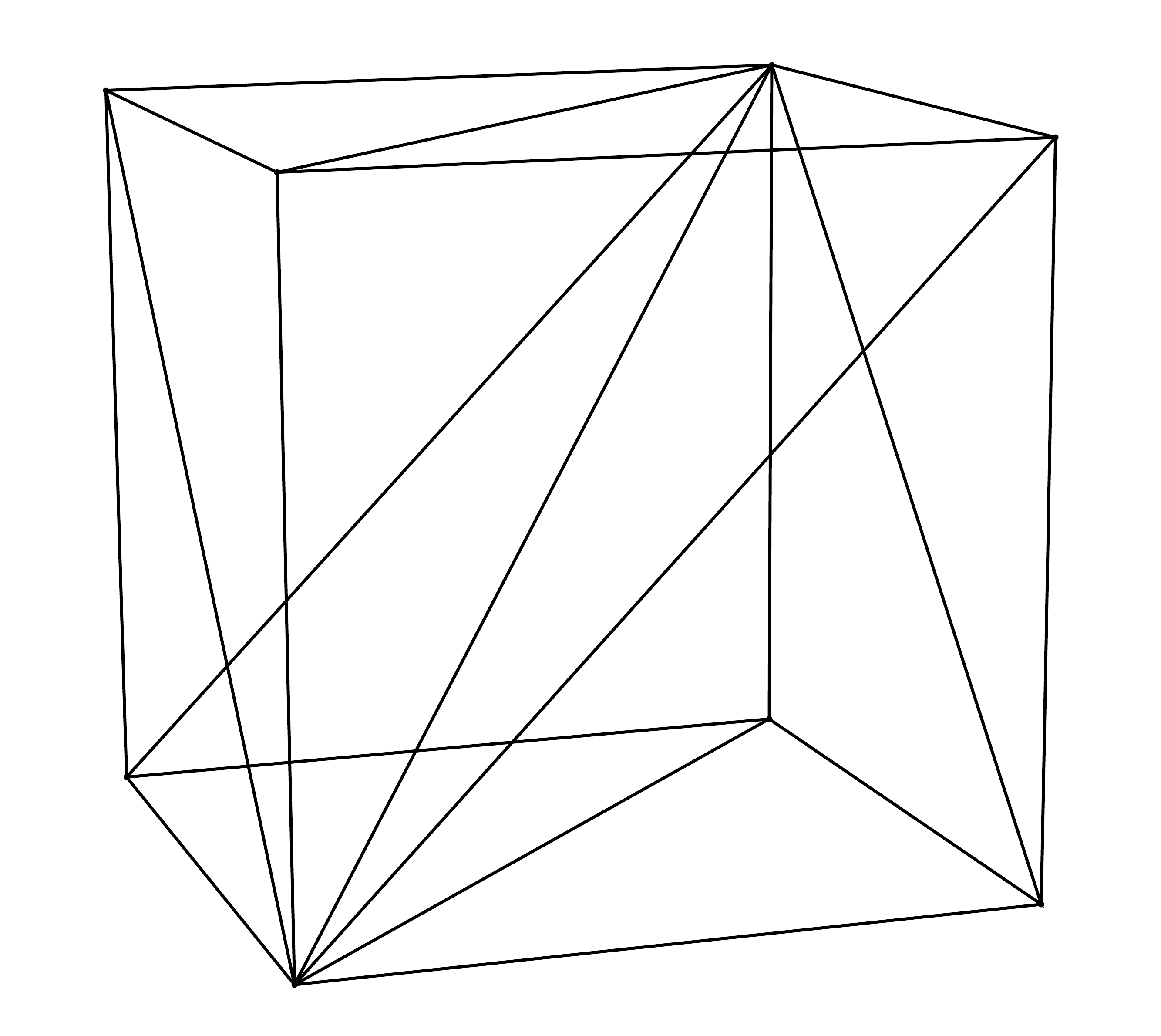}
\par\end{centering}

\caption{Elementary 3d mesh building block containing six tetrahedra, all of
which share the single interior diagonal.\label{fig:building-block}}

\end{figure}

In our computer calculations, we focus on pure gauge theory with $\mathcal{G}=SU(2)$,
and used a four-dimensional cubic euclidean domain $[0,1]^{4}\subset\mathbb{R}^{4}$
with periodic boundary conditions. The lattice structure consisted
of a three-dimensional simplicial mesh replicated at each discrete
time value. The three-dimensional simplicial mesh consisted of a homogeneous
arrangement of $N^{3}$ identical cubic building blocks, each building
block containing six tetrahedra as shown in figure \ref{fig:building-block}.
Each such spatial mesh was replicated $N$ times in the time direction
to fill the four-dimensional domain, in accordance with the construction
detailed in the previous sections. The lattice constant $h$ is defined
as the side length of each cubic building block, and also coincides
with the time discretization interval. In the interest of simplicity
we enforced temporal gauge, in which the temporal link matrices reduce
to indentity matrices.

The SGT action employs parallel transport matrices in order for gauge
invariance to be respected. By defining the distinguished points of
all spatial and temporal faces to coincide for as many pairs of faces
as possible, we only need the parallel transport matrices for terms
in the action involving pairs of temporal faces with no common nodes.

\begin{figure}
\centering{}\includegraphics[width=10cm]{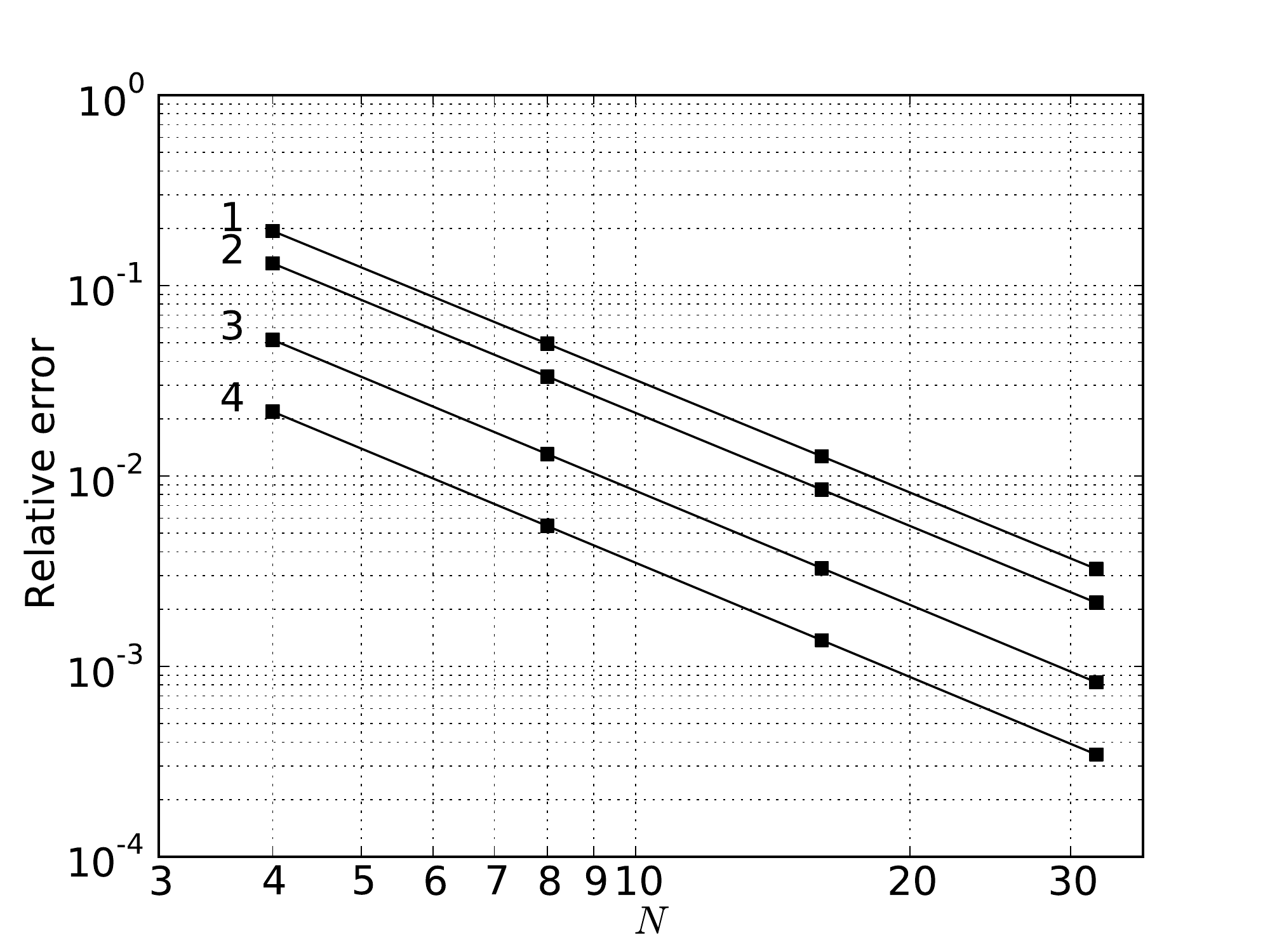}
\caption{The relative error of the action versus the number of lattice sites
per side $N$, for the actions 1, 2, 3, 4 described in section \ref{sec:num_convergence}.
The squares are the simulation data points and the solid lines are
the second order polynomial fits. Errors are proportional to $h^{2}$
in all cases.\label{fig:relative-errors}}
\end{figure}

In order to test convergence of the euclidean SGT, we compared the
discrete and continuum action for several different choices of gauge
fields for which the continuum action $S$ \ref{cont_action} can
be calculated exactly. We chose the following cases:
\begin{enumerate}
\item Gauge field oriented towards the $x$-direction in space and towards
the generator $t^{3}=i\sigma^{3}/2$ within $\mathfrak{su}(2)$, with
a sinusoidal time dependence, where $\sigma^{3}$ is a Pauli matrix.
The only nonzero component of the gauge field $A$ is 
\[
A_{x}^{3}(t,x,y,z):=\frac{1}{2\pi}\sin(2\pi t),\quad S=1.
\]
 
\item Gauge field oriented towards the $y$-direction in space and $t^{3}$
within $\mathfrak{su}(2)$, with a sinusoidal $x$-dependence. The
nonzero component of the gauge field in this case was 
\[
A_{y}^{3}(t,x,y,z):=\frac{1}{2\pi}\sin(2\pi x),\quad S=1.
\]
 
\item A case with two nonzero components, 
\[
A_{x}^{1}:=\frac{1}{2\pi}\sin(2\pi y),\quad A_{y}^{2}:=\frac{1}{2\pi}\sin(2\pi x),\quad S=\frac{1}{2}+\frac{1}{8(2\pi)^{4}}.
\]

\item A constant field that only contributes to the nonlinear term in the
field strength,
\[
A_{x}^{1}:=1,\quad A_{y}^{2}:=1,\quad S=\frac{1}{2}.
\]

\end{enumerate}
The first case is insensitive to the spatial face mass matrix elements,
while the second is insensitive to the spatial edge mass matrix elements
which are used in the definition of the temporal mass matrix elements.
In the first two cases, the nonlinear contribution to the continuum
field strength $F$ vanishes, and the action can be calculated analytically
to be unity. In the third case, the nonlinear term survives, and the
exact value of the action is 
\[
S=\frac{1}{2}+\frac{1}{8(2\pi)^{4}}.
\]
In all cases, we measure the relative error of the discretized action,
versus the lattice size, for lattices sizes $N^{4}$ from $N=4$ to
$N=32$. The results are displayed graphically in figure \ref{fig:relative-errors}. 

We plotted the absolute value of the relative deviation of the numerical
action from the continuum action, as a function of the number of lattice
sites per dimension. In all cases, the errors approach zero as the
lattice resolution grows, so the action converges towards the continuum
result. Using least squares second order polynomial fit, we determined
that the relative error depends in the following way on the lattice
constant $h$, 
\[
\mbox{Relative error}\approx Ch^{2},
\]
where $C$ is some constant depending on the choice of gauge field.
This is in accordance with the estimate \ref{eq:action-error}.

Quantum $SU(2)$ gauge theory Monte Carlo computer simulations using
the SGT action will be published in a companion article \cite{HalvSor2011b}.

\section{Conclusions}

We have proposed a general formulation of lattice gauge theory on
simplicial lattices. For any simplicial lattice of arbitrary shape,
this action can by used for lattice gauge theory simulations or study
of the classical equations of motion. Traditionally, lattice QCD simulations
within physics have used a homogeneous mesh. Mesh refinement is a
well established concept within the subject of FEM. We feel that it
is well worth the effort to invenstigate the possibility of mesh refinement
within classical and quantum gauge theory. Quantum lattice gauge theory
simulations are very computer intensive. Therefore mesh refinement
could be beneficial in cases where it makes sense to focus more computational
effort on some subset of the simulation domain.

We have shown the consistency of this SGT numerical approximation
to the continuous action, in the sense of approximation theory. The
lattice gauge theory formalism is of such a complexity, that it makes
sense to complement this theoretical proof with numerical {}``evidence''.
We have provided this for a few different cases of gauge fields, for
which the action was shown to converge towards the continuum result
as the grid fineness increased.

 \bibliographystyle{plain}
\bibliography{ref}

\begin{thebibliography}{10}

\bibitem{christiansen09}
Snorre~H. Christiansen and Tore~G. Halvorsen.
\newblock {A gauge invariant discretization on simplicial grids of the
  Schr\"odinger eigenvalue problem in an electromagnetic field}.
\newblock {\em E-print, UiO}, 2009.

\bibitem{SHC_TGH10}
Snorre~H. Christiansen and Tore~G. Halvorsen.
\newblock {A simplicial gauge theory}.
\newblock {\em ArXiv e-prints}, June 2010.

\bibitem{christiansen:75}
Snorre~H. Christiansen and Ragnar Winther.
\newblock {On Constraint Preservation in Numerical Simulations of Yang--Mills
  Equations}.
\newblock {\em SIAM Journal on Scientific Computing}, 28(1):75--101, 2006.

\bibitem{christiansen2011}
Snorre~Harald Christiansen, Hans~Z. Munthe-Kaas, and Brynjulf Owren.
\newblock {Topics in structure-preserving discretization}.
\newblock {\em Acta Numerica}, 20:1--119, 2011.

\bibitem{ciarlet78}
Philippe~G. Ciarlet.
\newblock {\em {The Finite Element Method for Elliptic Problems}}, volume~4 of
  {\em Studies in mathematics and its applications}.
\newblock North-Holland Publishing Company, 1. edition, 1978.

\bibitem{Creutz:1984mg}
Michael Creutz.
\newblock {\em {Quarks, Gluons and Lattices}}.
\newblock Cambridge, Uk: Univ. Pr. (Cambridge Monographs On Mathematical
  Physics), 1986.

\bibitem{hall04}
Brian~C. Hall.
\newblock {\em {Lie Groups, Lie Algebras, and Representations, An Elementary
  Introduction}}.
\newblock Springer, 2. edition, 2004.

\bibitem{HalvSor2011b}
Tore~G. Halvorsen and Torquil~M. S\o{}rensen.
\newblock {Lattice gauge theory using the Simplicial Gauge Theory action}.
\newblock {\em In preparation}, 2011.

\bibitem{hiptmair02}
Ralf Hiptmair.
\newblock {Finite elements in computational electromagnetism}.
\newblock {\em Acta Numerica}, 11(-1):237--339, 2002.

\bibitem{monk03}
Peter Monk.
\newblock {\em Finite Element Methods for Maxwell's Equations}.
\newblock Oxford Science Publications, reprinted edition, 2006.

\bibitem{munthe_owren99}
Hans Munthe-Kaas and Brynjulf Owren.
\newblock {Computations in a free Lie algebra}.
\newblock {\em Philosophical Transactions of the Royal Society of London.
  Series A: Mathematical, Physical and Engineering Sciences},
  357(1754):957--981, 1999.

\bibitem{Peskin:1995ev}
Michael~E. Peskin and Daniel~V. Schroeder.
\newblock {\em {An Introduction to quantum field theory}}.
\newblock Reading, USA: Addison-Wesley, 1995.

\bibitem{Weinberg:1995mt}
Steven Weinberg.
\newblock {\em {The Quantum theory of fields. Vol. 1: Foundations}}.
\newblock Cambridge, UK: Univ. Pr., 1995.

\bibitem{Weinberg:1996kr}
Steven Weinberg.
\newblock {\em {The quantum theory of fields. Vol. 2: Modern applications}}.
\newblock Cambridge, UK: Univ. Pr., 1996.

\bibitem{whitney57}
Hassler Whitney.
\newblock {\em Geometric integration theory}.
\newblock Princeton University Press, Princeton, N. J., 1957.

\bibitem{wilson74}
Kenneth~G. Wilson.
\newblock {Confinement of quarks}.
\newblock {\em Phys. Rev. D}, 10(8):2445--2459, Oct 1974.

\bibitem{Yang:1954ek}
Chen-Ning Yang and Robert~L. Mills.
\newblock {Conservation of isotopic spin and isotopic gauge invariance}.
\newblock {\em Phys. Rev.}, 96:191--195, 1954.

\end{thebibliography}

\end{document}